# Modular Assurance of Complex Systems Using Contract-Based Design Principles

**Dag McGeorge[1]* and Jon Arne Glomsrud[2]**

[1] Group Research & Development, DNV AS, Høvik, Norway
[2] Group Research & Development , DNV AS, Trondheim, Norway

*E-mail: dag.mcgeorge@dnv.com

**Abstract.** A growing number of safety-critical industries agree that building confidence in complex systems can be achieved through evidence and structured argumentation framed in assurance cases. Nevertheless, according to practical industry experience, assurance cases can easily become too rigorous and difficult to develop and maintain when applied to complex systems. Therefore, we propose to use contract-based development (CBD), a method to manage complexity originally developed in computer science, to simplify assurance cases by modularizing them. This paper will not only summarize relevant previous work such as constructing consistent modular assurance cases using CBD, but more importantly also propose a novel approach to integrate CBD with the argumentation in assurance case modules. This approach will allow subject-matter and domain experts to build assurance case modules together without having to know CBD. This can help a broader application of these methods in industry because subject matter experts outside of computer science can contribute to cross disciplinary co-development of assurance cases without having to learn CBD. Industry experience has proven four rules of thumb helpful for developing high-quality assurance cases. This article illustrates their usefulness and explains how modular assurance enables assurance that accounts for the interdependency of different concerns such as safety, security and performance.

## 1. Introduction

The goal of assuring a system is to justify confidence in properties of the system so that stakeholders can depend on it in their business and other activities. Assurance is defined in ISO/IEC/IEEE 15026 as 'grounds for justified confidence that a claim is or will be achieved' [1]. Claims express system properties of interest, and grounds for confidence in them are expressed as argumentation supported by evidence in an assurance case.

From the diverse arguments we use in everyday life and in our professions, one might speculate whether there are general rules or patterns of good argumentation at all. Clearly, assuring an autonomous navigation system of a ship is quite a different matter than assuring the ship's hull structure, a steel plate, a numerical simulation model or a chat bot. In his book *The uses of arguments* [2], the British philosopher Stephen Toulmin explored what characterizes good arguments and distinguished between the things about arguments that he called field-invariant and the things about them that he called field-dependent. What he called field-invariant has inspired some popular ways of expressing evidence-based arguments in assurance cases in a highly structured manner. Examples include the OMG Structured Assurance Case Metamodel (SACM) [3], Goal Structuring Notation (GSN) [4], Trust-IT [5], and Claim Argument Evidence (CAE) [6]. These methodologies, also referred to as evidence-based reasoning, combine logical reasoning, with the acknowledgement that our understanding of complex systems is often incomplete, and that the validity of claims needs to be evaluated in relation to the context the argumentation is used in.

The use of assurance cases in various forms is a long standing one, attracting considerably increased interest following the Piper Alpha disaster where explosions and fire on an oil platform caused 165 casualties. A conclusion in the public inquiry report was that "*The presentation of the formal safety assessment should take the form of a Safety Case, which would be updated at regular intervals and on the occurrence of a major change of circumstances*" [7]. Safety- and other assurance cases have later become common in diverse fields such as offshore technology [8], medical devices [9], systems and software engineering [1], road vehicles [10], and autonomous products [11], and a new generic recommended practice DNV-RP-0671, Assurance of AI-enabled systems [12], takes a modular approach and states "Assurance […] involves constructing an assurance argument consisting of claims […] substantiated by evidence".

For simple systems, comprehendible assurance case arguments can be expressed in the formats mentioned above. However, a challenge with assurance cases has been that they easily become overly rigorous and cluttered with nonessential detail when used in practical industry cases. A source of unnecessary detail is expressing other things than argumentation in the argument such as expressing as claims and argumentation the work processes that were used, project activity plans, models of the system itself and its architecture, the methods or theories used to analyze the system and create evidence etc. For all these things, tools more suitable than argumentation often exist such as flow diagrams, project management tools, system models, recommended practices etc. Hence, to avoid unnecessary rigor and clutter, it is recommended to:

**Rule of thumb 1:** Use arguments for argumentation only.

Of particular interest here is how to account for the methods and approaches selected to generate evidence for the assurance case, such as e.g. system models, test methods, simulations, hazard analyses etc. As alluded to above, it is possible to express them as claims and argumentation, however when well established in their fields, and where tailored guidelines, practices and tools exist, these should be used, referenced in the assurance case, and the results that they provide should be used as evidence supporting the assurance case. Such a practice is a way of respecting Rule of thumb 1 and can help reducing clutter and nonessential detail in the assurance case.

Extending onto this, it is recommended to reserve the assurance argumentation for connecting the various established methods and approaches rather than capturing their details in the argumentation. This perspective was also taken by Bloomfield and Rushby [13]. They use the term *theory* to denote the methods and approaches and suggest structuring the argument around the selected theories. That terminology is adopted here. As Bloomfield and Rushby put it: "When using a theory, the argument must provide justification that it is suitable and credible, and that it is applied appropriately but it does not present the theory as part of the argument: it merely references it" [13]. Hence the recommendation to:

**Rule of thumb 2:** Leverage established theories to structure the argumentation.

In industrial cases, involving many different subject matter fields of expertise, many contributions from different parties are needed to realize and operate the system. Capturing this complexity in argumentation would lead to very complicated and interlinked arguments. Modularizing assurance cases can help managing this kind of complexity. Hence the recommendation to follow:

**Rule of thumb 3:** Encapsulate complexity in modules that can be assured separately.

Nešic et al [14] have already published an approach to combining contract-based design (CBD) and modular assurance cases. Their approach to contract-based design is adopted here almost without change, while an attempt is made at simplifying and further developing the way of connecting CBD to modular assurance case argumentation. The benefit of the new approach stems from combining the formal rigor in assuring consistency of specifications across the system levels with confining field-dependent evidence-based argumentation inside modules that can be dealt with by subject matter experts unfamiliar with CBD.

This paper aims to share practical industry experience and is organized as follows:
- Section 2 summarizes relevant previous work on assurance cases (Section 2.1) and contract-based design (CBD) (Section 2.2).
- Section 3 presents our novel approach to modular assurance.
- Section 4 discusses the differences and advantages of the proposed approach.
- The main conclusions are summarized in Section 5.

## 2 Summary of previous work

*2.1 Assurance Cases*

The function of an assurance case is to substantiate a claim by evidence and argumentation. ISO/IEC 15026 defines inference as a reasoning step that derives a claim from a list of sub-claims (premises) under a specified context. The claim derived by the inference is the conclusion. The claims from which the inference derives the conclusion are the premises. Confidence in the conclusion depends on the validity of the inference, which is expressed as a justification, and the confidence in the premises.

A graphical representation called goal structuring notation [4] is used here to show assurance case argumentation, with some adjustments explained in Section 4.1. Figure 1 illustrates a general GSN argumentation pattern. Rectangles represent claims, both the premises and the conclusion. The context element describes the context in which the validity of argumentation shall be judged.

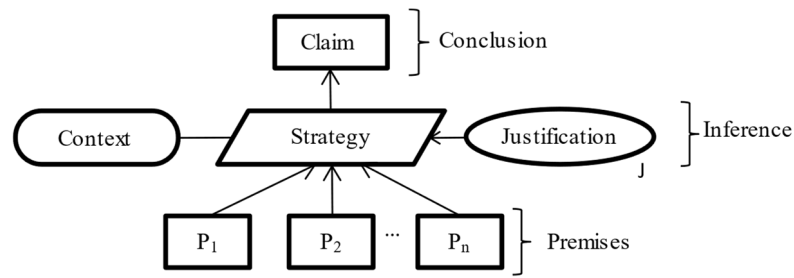

**Figure 1.** Format used to show assurance case argumentation

Practical experience with assurance case development suggests that:
1. When a claim of interest has been identified, one would search for a way to support the claim in the given context – a good assurance strategy – rendered as a parallelogram. That could e.g. be to test something in a laboratory, simulate it on a computer or consult subject matter experts. This is a choice among options. The term 'strategy' alludes to the optionality in top-down construction of arguments.
2. The validity of a strategy in a given context is hardly obvious. The justification, rendered as an oval with the letter 'J', should capture explicitly the reasons for considering the strategy valid in the circumstances.
3. It follows from the above that confidence in the concluding claim depends on the confidence in the justification and the confidence in each of the premises, including their supporting argumentation and evidence. Without a stated justification, the validity of the conclusion remains undefined. Justified confidence in all these elements automatically implies confidence in the conclusion. This removes the need for a special 'confidence argument'.

The validity of the argument depends on the context of the argumentation. Context information should include descriptions and models of the system, the environment in which the system is deployed and the way that it is intended to be used.

When Rule of thumb 2 (leverage established theories) is applied:
- The argumentation strategy would be to apply the selected theory.
- The justification explains why it is acceptable to use that theory in the circumstances.
- The results from application of the theory would be the supporting evidence.
- The conclusion (claim) would be the one justified according to the theory.

*2.2 Contract-Based Design (CBD)*

The systematic use of contracts in design was introduced by Benveniste et al, see e.g. [15] for a recent account of their work. The use of modules connected via contracts has also become a recommended practice for assurance of simulation models [16].

This work is based on the work of Nešic et al [14] which provides a detailed description of contract-based design (CBD) with formal proofs. CBD provides a way to ensure consistency of the specifications across the system. The interested reader is referred to [14] for details. The description here is kept brief and simple, with the aim of making it accessible to the engineering community while explaining how CBD maps to assurance cases in a way that is scientifically justified. Together with the approach of [17], which assures completeness of the contracts including component interaction and emergent behaviour, this can be used to follow Rule of thumb 3 (encapsulate complexity in modules).

At the core of CBD is the concept of *assume-guarantee contracts*, defined as an ordered pair of specifications where the first specification is a finite set of *assumptions* and the second specification is a *guarantee*. Guarantees are used to express properties that a component should provide. The other components that it interacts with, and the external world, are referred to as the environment, which implements the assumptions. A component $C$ that satisfies a contract $K = (A_i, G)$ can be developed independently of other components. The component can be said to implement the contract. Component $C$ considered together with its environment $C_e$ (which implements the assumptions in the contract $K$) implements the guarantee $G$.

A slightly modified version of the example of Nešic et al [14] is used to explain the key concepts of CBD. Figure 2 shows the CBD specification structure. It contains the components of a system, their relations and all the contracts allocated to the system and its components. There

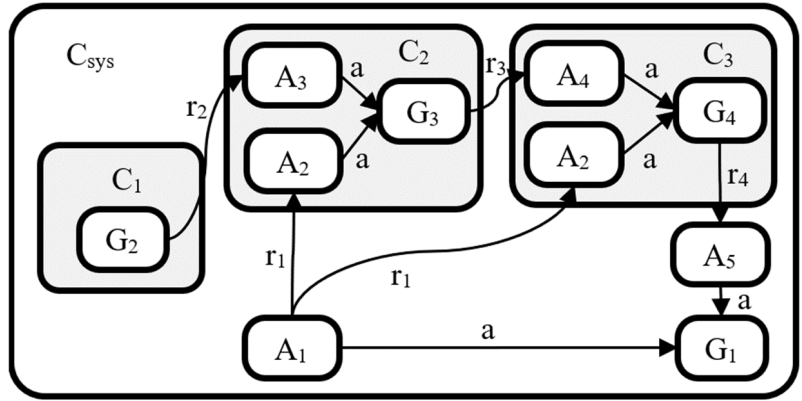

**Figure 2.** The specification structure of system $C_{sys}$ (Fig 5.3a in Nešic et al [14] slightly adjusted)

are two kinds of relationships:

1. *Assumption-of relations* (marked with a) between an assumption and a guarantee.
2. *Refinement-of relations* (marked with r) between specifications of the components.

The assumption-of relations mapped onto a component show the contracts allocated to that component. For example: $C_{sys}$ has allocated $K_{sys} = (\{A_1, A_5\}, G_1)$, i.e. if $A_1$ and $A_5$ are satisfied, then $C_{sys}$ guarantees $G_1$.

The refinement-of relations can be understood from the concept of contract refinement. If a contract $K'$ is a refinement of a contract $K$, it means that a more capable component is needed to satisfy the refined contract $K'$ than to satisfy contract $K$. Being more capable means that the component imposes less constraints on its environment or that it offers stronger guarantees. The refinements specify dependencies:

1. a component that depends on an assumption made by the system it is part of,
2. a component that depends on the guarantees of siblings within the same system, or
3. a system that depends on the guarantee of a component within it.

The refine relationships always go from the independent system or component to the one that depends on it. The specification structure (Figure 2) shows for example that:

1. $A_1$ is a refinement of $A_2$, i.e if $A_1$ is satisfied one can take it that $A_2$ is too.
2. $G_2$ is a refinement of $A_3$, i.e. if $G_2$ is satisfied we can take it that $A_3$ is too.
3. $G_4$ is a refinement of $A_5$, i.e. if $G_4$ is satisfied we can take it that $A_5$ is too.

## 3 Modular assurance

*3.1 Assurance of the System*

**Objective.** The objective of assuring $C_{sys}$ is to provide confidence that $C_{sys}$ satisfies its allocated contracts. In the example there is just one contract $K_{sys}$ allocated to $C_{sys}$ with the guarantee $G_1$. Hence, what is needed is a modular assurance case that justifies confidence in $G_1$. If there were more than one guarantee allocated to $C_{sys}$, one would have to justify confidence in each one.

**Using CBD as a Theory.** In the spirit of Rule of thumb 2, CBD is adopted as a theory. Hence it is taken as preconditions for the system assurance that a CBD analysis of the system has been performed, and that the specification structure of the system (Figure 2) has been documented as an assurance artifact that can be used as evidence. Furthermore, the assurance artifact is taken to have been verified by competent subject matter experts and therefore can be trusted as evidence. This simplifies the assurance case considerably because the parts of the argumentation showing correct use of CBD is confined within the evidence in the system integration assurance module ($C_{sys}$ in the example) and need not be captured by the assurance modules of the component and refinement shown in the GSN diagrams. This latter is an application of Rule of thumb 1 (use arguments for argumentation only).

**Defining the Assurance Case Module Architecture.** The assurance case must cover the system integration, which in the CBD specification structure is represented by the refinement relations. We use the fourth rule of thumb for this purpose, which allows to simplify the assurance argumentation as will become apparent later:

**Rule of thumb 4:** Reflect the system architecture in the assurance case architecture.

This suggests using the following assurance modules shown in Figure 3:
1. Four assurance modules showing that the components satisfy their allocated contracts, one for the system $C_{sys}$ and one for each component $C_1$, $C_2$, and $C_3$
2. Four refinement modules showing that the refinements between the components are valid, one for each of the four refinement relations $r_1$ to $r_4$.

GSN architecture view notation [4] is used in Figure 3 with some adjustments explained in Section 4.1. GSN modules represent the components, *GSN contracts* represent the refinements and arrows showing the bindings between the assurance modules (Section 1.4.4 in [4], see particularly Fig 1:4-7). This reveals the one-to-one mapping from the components and refinements in the specification structure (Figure 2) to the assurance module architecture (Figure 3). The term *GSN contract* is used here for consistency with the GSN standard and should not be confused with the CBD concept of *assume-guarantee contracts* described in Section 2.2.

Capturing the system architecture and relations between components in the assurance case architecture removes the need to capture them in the argumentation. This is another application of Rule of thumb 1. And it is the key to simplifying the assurance modules as shown in the following sections, which illustrates the benefit of using Rule of thumb 4 (reflect the system architecture in the assurance case architecture).

The scope for the system assurance can now be clarified as making assurance case modules for each component, see Section 3.2, and each refinement, see Section 3.3.

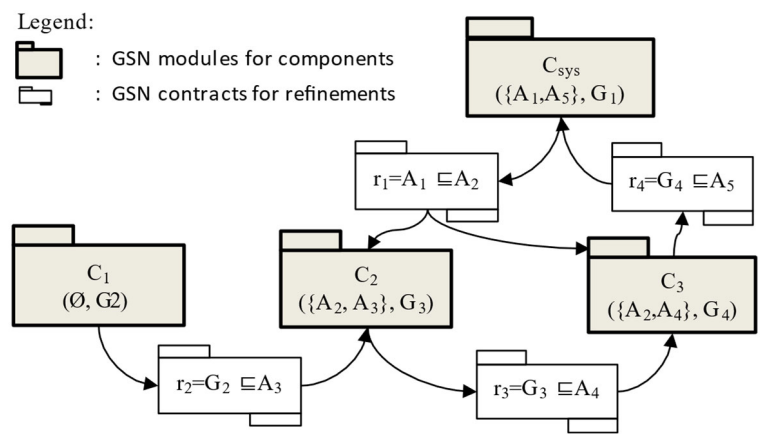

**Figure 3.** Assurance module architecture.

*3.2 Assurance Modules for Components*

The component assurance modules must substantiate that the guarantee $G$ of each allocated contract is justified provided that the assumptions $A_j$ about the component's environment are respected. The environment consists of everything outside of the component, which includes all the other components including any sub-components. This is captured by the *a*-relationships in the specification structure in Figure 2.

Hence, an argument is needed that leads from the contract's assumptions as premises to the contract's guarantee as conclusion. A basic pattern serving this purpose is illustrated in Figure 4. The direction of the assumption-of relationships in Figure 2 corresponds with the direction of argumentation from premises to conclusion in Figure 4.

The field-dependent argumentation substantiating $G$ from $A_i$, shown grey between the horizontal scope lines in Figure 4, can involve many strategies and layers of argumentation and component-specific evidence. CBD does not provide insight into the contents or format of these field-dependent arguments. The contract just specifies what the subject matter experts need to provide argument and evidence for.

If several contracts are allocated to a component C, the template can be reused.

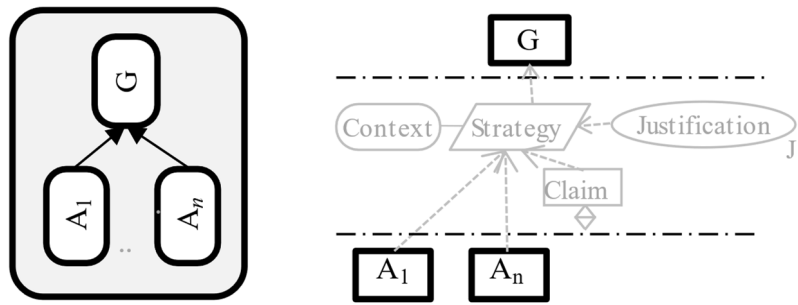

**Figure 4.** Assumption-of relationships (left) and the corresponding argument template for assurance modules for components (right).

*3.3 Assurance Modules for Refinements*

The refinement modules are represented as GSN contracts, which are different from CBD assume-guarantee contracts. To avoid confusion, these assurance modules are referred to here as refinements modules. To provide intuition and simplify comprehension, the notion of dependency between components is used:

1. $C_{sys}$ depends on $C_3$, because if $C_3$ fails to provide $G_4$, $C_{sys}$ would fail to provide $G_1$.
2. $C_3$ depends on $C_2$ which depends on $C_1$, due to the chain of refinements.
3. $C_2$ depends on $C_{sys}$ because $C_2$'s guarantee depends on an assumption $A_1$ of $C_{sys}$.

To explain how the refine relations map to argumentation, consider a refine-relation that goes from an assumption or guarantee in component $C_i$ to assumption $A$ in component $C_j$. The refinement and its representation as a refinement assurance module connecting two component modules $C_i$ and $C_j$ are illustrated in Figure 5 (left and middle). If the components are assured already (they satisfy their contracts), what remains to show is that the assumption of $C_j$ follows logically from assumption or guarantee ($A/G$) of $C_i$. This can be done by an argument leading from $A/G$ of $C_i$ as a premise to the assumption of $C_j$ as a conclusion. An argumentation template for such a refinement module is illustrated in Figure 5 (right). Intuitively, the direction of the refinement relationship corresponds with the direction of argumentation from premises to conclusion.

The assurance modules for components and refinements can now be developed and verified independently. The binding between them is defined by the CBD specification structure (exemplified in Figure 2) and shown in the assurance module architecture (exemplified in Figure 3). According to the CBD theory, this verifies the components and their integration, and is enough to justify confidence in the system guarantees.

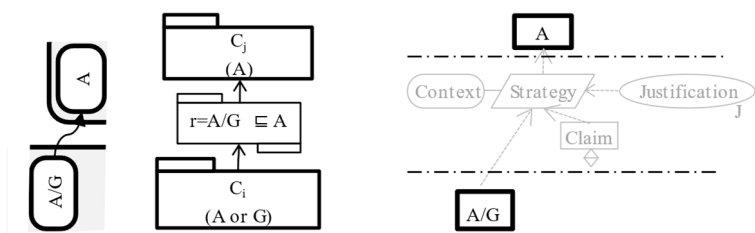

**Figure 5.** CBD refinement relation (left), corresponding fragment of assurance module architecture (middle) and argument template for assurance modules for refinements (right).

## 4 Discussion

*4.1 Relations to previous work*

**Slight Deviations from the GSN Standard.** The names of the nodes in the GSN diagrams have been adapted to the conventions of ISO/IEC 15026 [1]. For example, the term claim is used instead of goal. The arrows in the argumentation and architecture diagrams are reversed compared to the GSN standard [4], showing the direction of reasoning from premises to conclusion. Then, intuitively, the direction of the arrows in the argumentation and architecture diagrams coincide with the directions of the assumption-of and refinement relationships in the CBD specification structure.

**A Slight Adjustment to the CBD Theory.** Nešic et al [14] allow goal-to-goal refinement (e.g. $G_4 \sqsubseteq G_1$ in their example). Here an additional assumption $A_5$ was introduced between $G_4$ and $G_1$ to restrict the refinements to the forms $A_j \sqsubseteq A_i$ and $G_j \sqsubseteq A_i$ (guarantees or assumptions are refinement of assumptions). This was done to align with the goal of modular assurance to allow all systems, subsystems and components to be assured without knowledge about other systems, subsystems and components. In the example, $C_{sys}$ does not need to know about $C_3$, it merely assumes $A_5$. As a separate step in the system integration of $C_{sys}$, the refinement from $G_4$ to $A_5$ is assured.

A benefit of this arises when the modular assurance case is derived from the specification structure. With goal-to-goal refinement, the component assurance module for $C_{sys}$ in the example needs to combine the argumentation for contract satisfaction (from $A_1$ to $G_1$) with refinement argumentation (from $G_4$ to $G_1$), because $G_1$ not only depends on contract satisfaction but also on a refinement. If component $C_3$ is replaced, so that $G_4$ is adjusted, the entire argumentation for $C_{sys}$ would have to be scrutinized. With the added assumption $A_5$, the component assurance module for $C_{sys}$ would remain unchanged, confining any changes to the refinement assurance module.

**Field-invariant and Field-dependent Aspects of Modular Assurance.** Toulmin [2] identified field-invariant aspects of arguments. This concept has been extended here to modular arguments. The translation of the CBD contract specification to an assurance module architecture (Section 3.1) and assurance case module templates (Sections 3.2 and 3.3) constitute the field-invariant aspects of modular assurance.

CBD does not provide insight into the behaviour of the system. That requires expertise in the relevant subject matter fields to capture subsystem behaviour and interactions between subsystems. These field-dependent aspects of modular assurance provide:
1. The verification of the sufficiency (completeness) of the contracts.
2. Argumentation and evidence generation inside the *component* assurance modules.
3. Argumentation and evidence generation inside the *refinement* assurance modules.

**Comparison with the Assurance Case Argument Patterns of Nešic et al.** Nešic et al [14] provide three argumentation patterns in GSN format. These patterns mirror the principles of the CBD theory. To represent the complexity of CBD they make use of advanced GSN argument pattern and modular extensions (section 1:3 and 1:4 of [4]), and refer to the theorems, corollaries and definitions of CBD. The expressions used in the argumentation also use the concepts and terminology of CBD. Practical use of these templates for any component or system thus requires knowing the CBD theory and advanced GSN extensions. That may present a severe obstacle to adoption in industry, where practicing engineers are experts in their subject matter fields rather than experts in CBD and GSN.

The alternative approach advocated here uses CBD to derive the assurance case architecture. This requires knowledge of CBD. However, the outcomes of the CBD analysis are simple module templates that capture assumptions and guarantees about the properties of systems and components, which should be familiar to the subject matter experts assigned to develop the field-dependent argumentation in the assur-ance case modules.

*4.2 Scalability, Reuse and Efficiency*

**Scalability.** Cross-disciplinary work is required to assure complex systems, but it is also an obstacle to adoption and scaling because it requires tedious knowledge sharing and learning. The proposed approach, combining CBD and assurance cases, separates concerns and reduces waste from unnecessary cross-disciplinary coordination. Instead of requiring all subject matter experts to know CBD and advanced GSN extensions, knowledge of CBD is used only to establish the assurance case architecture and only simple GSN concepts are used. Then, subject matter experts can team up to solve the required challenges within their fields.

**Reuse of Assurance Case Modules.** The connections from an assurance case module to its surroundings are captured by the assumptions in its allocated contracts. A change to the system, its environment or its use, may or may not affect the assumptions of a particular assurance module. If the assumptions of a module remain unchanged, it can be reused without further verification or development. Similarly, the refinement modules can be reused as long as the refinements remain unchanged. In this way, changes are confined within modules, minimizing rework and maximizing reuse.

**Product Line Assurance.** A key contribution in the paper by Nešic et al was to address assurance of product lines [14]. Product line engineering facilitates the development of a family of systems that are jointly referred to as a product line. Instead of assuring each system variant individually, the product line approach allows assuring each component featuring in a variant and their integration into the system only once, and then providing assurance of specific system variants whenever needed simply by combining the assurance of individual components. This is a huge advantage whenever the number of relevant combinations is high compared to the number of components they consist of.

Extending the present work to product lines is straightforward. The methodology of Nešic et al [14] can be used to construct the CBD specification structure. Then assurance modules can be established for each component and refinement needed for the variants of interest. For a particular variant, the assurance module architecture (exemplified in Figure 3) is defined by the specification structure. The modular assurance case for that variant is established just by combining the assurance modules for the relevant components and refinements as described by Nešic et al.

**Multi-Concern Assurance.** Typical system properties of concern, such as safety, security and performance emerge on the system level from the components' properties and interactions. The components' properties are captured in the component contracts and the interactions are captured in the refinements, both of which can be relevant for different concerns and reused as need be. The assurance case module architecture captures how modules contribute to assuring the different concerns. This simplifies the assurance argumentation because it reduces the needs to cross-link between the argumentations for the different concerns.

**Automation of Assurance Case Construction.** The straightforward translation from CBD to assurance module templates lends itself to automation, as the translation does not require knowledge of the system or expert judgements. These are the field-invariant aspects of modular assurance. The module templates would then have to be completed with field-dependent argumentation that captures the relevant subject matter knowledge. That is harder to automate. New large language models could provide automatic aids to experts charged to develop the field-dependent argumentation.

## 5 Conclusions

This work attempts making the benefits of CBD and modular assurance more accessible to practicing engineers in industry. The proposed novel approach does this by:
- Dividing the assurance case into modules in a way that shields the subject matter experts from the probably unfamiliar CBD theory.
- Using experts to verify component and system properties, which tend to be their field of expertise, rather than overwhelming them with unfamiliar abstract theories.

This aims to help scaling of this methodology and offer efficiency and quality gains to industry. Modular assurance cases can also simplify multi-concern assurance because assurance case modules can be reused across the different concerns.

The usefulness of some simple rules of thumb was also illustrated, that can help practitioners making high quality modular assurance cases:
1. Use arguments for argumentation only.
2. Leverage established theories to structure the argumentation.
3. Encapsulate complexity in modules, which can be assured separately.
4. Reflect the system architecture in the assurance case architecture.

Another important contribution was the distinction between field-invariant parts of modular assurance, that lend themselves to tool support and automation, and the field-dependent parts that require subject matter expertise.